\documentclass{iopart}

\usepackage{iopams}
\usepackage{psfrag}
\usepackage{graphicx}
\usepackage{bm}% bold math
\usepackage{sidecap}

\newcommand{\ket}[1]{\left|#1\right\rangle}
\newcommand{\bra}[1]{\left\langle #1\right|}
\newcommand{\bracket}[1]{\left\langle #1\right\rangle}

\begin{document}

\title[The two-spinon transverse structure factor of the gapped Heisenberg AFM chain]{The two-spinon transverse structure factor of the gapped Heisenberg antiferromagnetic chain}

\author{Jean-S\'ebastien Caux and Jorn Mossel}
\address{Institute for Theoretical Physics, University of Amsterdam, Valckenierstraat 65, 1018 XE Amsterdam, The Netherlands.}
\author{Isaac P\'erez Castillo}
\address{Department of Mathematics, King's College London, Strand, London WC2R 2LS, United Kingdom}

\date{\today}

\begin{abstract}
We consider the transverse dynamical structure factor of the anisotropic Heisenberg spin-$1/2$ chain (XXZ model) 
in the gapped antiferromagnetic regime ($\Delta > 1$).  Specializing to the case of zero field, we use two 
independent approaches based on integrability (one valid for finite size, the other for the infinite lattice) 
to obtain the exact two-spinon part of this correlator. 
We discuss in particular its asymmetry with respect to the $\pi/2$ momentum line, its overall anisotropy dependence,
and its contribution to sum rules.
\end{abstract}

\maketitle

\section{Introduction}
Since the early days of quantum mechanics, Heisenberg spin chains \cite{HeisenbergZP49}
have been one of the most fertile laboratories for investigating strong quantum effects in
physically realistic condensed matter systems.  Perhaps the most important conceptual 
development associated to spin chains is Hans Bethe's famous Ansatz \cite{BetheZP71} for
their wavefunctions, leading in particular to exact thermodynamics for these and a wide variety of other types of 
integrable models (see \cite{KorepinBOOK,TakahashiBOOK} and references therein).

Another extremely important concept which studies of spin chains help to pin down is that of
particle transmutation.  Namely, when correlations become important, the microscopic degrees 
of freedom transform into effective quasi-particles barely resembling those of a free or weakly-coupled system.
For Heisenberg spin chains, this occurs when spin waves (which are the exact excitations at the
saturation field) give way to quantum solitonic excitations known as spinons \cite{FaddeevPLA85} 
(which are the exact excitations in zero field).  The power of integrability is illustrated by the fact 
that it allows this whole transmutation process to be described accurately.
Since spinons should be understood as excitations created over the highly entangled ground state of the spin chain, 
they possess a lot of richness, at
the simplest level in their dispersion relation, but also at a more elaborate level in their dynamics
(in other words their ability to `carry' correlations around in the system).  The latter point is quantified by their
contribution to various correlation functions, most important of which are the dynamical spin-spin
correlation functions which determine inelastic neutron scattering amplitudes.  
Various guises of integrability can now be used to address the question of correlation functions
in such strongly-correlated systems, and the purpose of this paper is to investigate such a case.

Spinons come most radically to the forefront for the anisotropic Heisenberg model
\begin{eqnarray}
H = J \sum_{j = 1}^N \left[ S^x_j S^x_{j+1} + S^y_j S^y_{j+1} + \Delta S^z_j S^z_{j+1}\right]
\label{XXZ}
\end{eqnarray}
in the gapped antiferromagnetic region $\Delta > 1$.  
For infinite $\Delta$, excitations are given by localized domain walls \cite{VillainPB75}
which can be seen as solitons of unit length in the lattice spacing.  States involving pairs of
such solitons contribute to neutron scattering amplitudes \cite{IshimuraPTP63} in the vicinity of this anisotropy limit.
In the region $1 < \Delta < \infty$, these solitons acquire a finite extent, and their structure and
behaviour requires methods such as integrability to be properly understood.  Experimental realizations of
spin chains in this regime include $\mbox{CsCoCl}_3$ \cite{YoshizawaPRB23,GoffPRB52} and
$\mbox{CsCoBr}_3$ \cite{NaglerPRL49,NaglerPRB28}.  An interesting fact is that there exists a clear separation of
energy scales between states with $2, 4$ and higher spinon numbers, which provides a way to isolate these
separate subclasses of states in measurable response functions.  We will in what follows be 
interested in the transverse dynamical structure factor (TDSF), defined as
\begin{eqnarray} 
S^{-+}(k, \omega) = \frac{1}{N} \sum_{j,j'=1}^N e^{-ik (j - j')} \int_{-\infty}^{\infty} dt e^{i\omega t}
\langle S^-_j (t) S^+_{j'} (0) \rangle,
\label{TDSF}
\end{eqnarray}
where the angular brackets $\langle ... \rangle$ denote a zero-temperature average (ground-state
expectation value).  In zero field, which is the case that we will be concerned with, we have
$S^{-+}(k, \omega) = S^{+-} (k,\omega) = 2S^{xx}(k,\omega) = 2S^{yy}(k,\omega)$.  
Our purpose will be to compute the contribution to this quantity coming from
all two-spinon intermediate states, using two separate methods relying on integrability.  The first
method applies to finite lattices, and makes use of determinant expressions for spin operator form
factors derived within the Algebraic Bethe Ansatz \cite{KitanineNPB554,KitanineNPB567} and 
used to compute structure factors of Heisenberg chains for general fields and anisotropies,
both for two-particle \cite{BiegelEL59,BiegelJPA36,SatoJPSJ73} and general multiparticle states
\cite{CauxPRL95,CauxJSTATP09003}.  The second method starts from an algebraic analysis
of the infinite chain in zero field \cite{JimboBOOK}, and uses the quantum group symmetry of the model to express
states and form factors directly in the thermodynamic limit.  
This method was first used to obtain the two-spinon part of the structure factor for the
isotropic (XXX) antiferromagnet in zero field \cite{BougourziPRB54,KarbachPRB55}, and can also
give the four-spinon part \cite{AbadaNPB497,CauxJSTATP12013}.  The two-spinon part of the $\Delta > 1$
regime was dealt with extensively in an earlier
paper \cite{BougourziPRB57}, which we would here like to revisit with a number of observations.

The paper is organized as follows.  We treat the finite and infinite chains in parallel, 
beginning with a characterization of eigenstates, emphasizing on the two-spinon states
which will be used to perform the partial trace over intermediate states when calculating the structure factor.  
Section \ref{sec:TDSF} is then devoted to the step-by-step construction of the TDSF, gluing together the
necessary dynamical constraints and form factors.  Our results are then presented and discussed, making the 
correspondence between the finite and infinite chains explicit.  Sum rules and anisotropy effects are
then discussed, and we end by offering some perspectives on possible future developments.

\section{Eigenstates}
\label{sec:Eigenstates}
\subsection{Finite lattice  Bethe Ansatz}
In the gapped antiferromagnetic sector $\Delta > 1$, the eigenstates of (\ref{XXZ}) are Bethe wavefunctions
which are each individually characterized by a set of rapidities $\{ \lambda_j \}$
solving the Bethe equations
\begin{equation}
\theta_1 (\lambda_j) - \frac{1}{N} \sum_{k=1}^M \theta_2 (\lambda_j - \lambda_k) = 2\pi \frac{I_j}{N}.
\label{BAE_XXZ_gpd_log}
\end{equation}
Here, $N$ is the number of sites (which we take to be even) 
and $M$ is the number of reversed spins starting from the ferromagnetic reference state with all spins up
(the magnetization is thus $S^z_{tot} = N/2 - M$).  
$I_j$ are quantum numbers (half-odd integers for $M$ even, integers for $M$ odd) which 
uniquely specify a set of rapidities (and thus an eigenstate) through the Bethe equations.
The kernels appearing in the Bethe equations are given by
\begin{equation}
\theta_n (\lambda) \equiv 2~\mbox{atan} \left[ \frac{\tan(\lambda)}{\tanh(n\eta/2)} \right] +
2\pi \left \lfloor \frac{\lambda}{\pi} + \frac{1}{2} \right \rfloor.
%\!\!-\! \frac{1}{N} \!\sum_{k = 1}^M 
%2\mbox{atan} \!\!\left[\frac{\tan(\lambda_j - \lambda_k)}{\tanh \eta}\right] \!\!=\! 2\pi \frac{I_j}{N}.
\end{equation}
The integer part of the kernel (second term) guarantees monotonicity for real $\lambda$.  This is important
for the classification of states, which we will discuss below.

The energy of a state in an external magnetic field $h$ is given
as a function of the rapidities by
\begin{eqnarray}
E = J \sum_{j = 1}^M \frac{-\sinh^2 \eta}{\cosh \eta - \cos 2\lambda_j} - h(\frac{N}{2} - M), 
\end{eqnarray}
whereas the momentum, which is the sum over quasi-momenta associated to each rapidity, 
has a simple representation in terms of the quantum numbers:
\begin{eqnarray}
K = \sum_{j = 1}^M \frac{1}{i} \ln \left[\frac{\sin(\lambda_j + i\eta/2)}{\sin(\lambda_j - i\eta/2)}\right]
= \pi M - \frac{2\pi}{N}\sum_{j = 1}^M I_j \hspace{0.5cm} \mbox{mod} \hspace{0.2cm}2\pi.
\label{Ep_XXZ_gpd}
\end{eqnarray}
The ground state is given by $I_j^0 = -\frac{M+1}{2} + j$, $j = 1, ...,M$, and is not degenerate.
However, there exists a quasi-degenerate
state with momentum $\pi$, given by adding an Umklapp to the true ground state.  This state becomes exactly
degenerate with the true ground state in the thermodynamic limit, whereas all other excited states remain gapped by a
finite value.  

Excited states are constructed by modifying this choice of quantum numbers.  In everything that follows,
we concentrate on solutions to the Bethe equations (\ref{BAE_XXZ_gpd_log}) taking the form of sets of
real rapidities only, $\{ \lambda_j \} $ with $\lambda_j \in \mathbb{R} ~\forall j$.  We thus do not 
consider string states;  however, the important two-spinon states fall into the class of states we consider.

Due to the monotonicity of the kernels $\theta_n$, the left-hand side of (\ref{BAE_XXZ_gpd_log}) is also
monotonic in $\lambda_j$.  We therefore have $\lambda_j < \lambda_k$ iff $I_j < I_k$.  Equal rapidities do
not yield proper Bethe wavefunctions, and we therefore only need to consider sets of distinct quantum numbers.
Moreover, in view of the parametrization of momenta in terms of rapidities used in (\ref{Ep_XXZ_gpd}), we
can restrict to solutions of the Bethe equations in terms of ordered sets of rapidities within an interval
of width $\pi$, {\it i.e.} such that $\lambda_M < \lambda_1 + \pi$.  Substituting this condition in the
difference of the Bethe equations for $\lambda_M$ and $\lambda_1$ gives a constraint on $I_M - I_1$, namely
\begin{equation}
\fl
\frac{2\pi}{N} (I_M - I_1) = 2\pi - \frac{2\pi}{N} \sum_{l=1}^M \left( \left \lfloor \frac{\lambda_M - \lambda_1}{\pi} + \frac{1}{2} \right\rfloor
- \left \lfloor \frac{\lambda_l - \lambda_1}{\pi} + \frac{1}{2} \right\rfloor \right) < 2\pi (1 - \frac{M}{N})
\end{equation}
so $I_M - I_1 < N - M$ iff $\lambda_M - \lambda_1 < \pi$.  Denote as ${\cal C}_M$ the set of all ordered sets of quantum
numbers such that $I_M - I_1 < N - M$.  Since $\lambda$ is identified with $\lambda + \pi$ as far as the wavefunctions are concerned, 
this set does not lead to a one-to-one mapping of quantum numbers with wavefunctions.
We can define a transformation $S: \{ ( \lambda_j, I_j ) \} \rightarrow \{ ( \tilde \lambda_j, \tilde I_j ) \}$ 
with $( \tilde \lambda_j, \tilde I_j ) = ( \lambda_{j+1}, I_{j+1} + 1 )$ for $j = 1, ..., M-1$ and
$( \tilde \lambda_M, \tilde I_M) = ( \lambda_1 + \pi, I_1 + N - M + 1 )$.  Then, $\tilde \lambda_j < \tilde \lambda_k$
iff $\tilde I_j < \tilde I_k ~ \forall j, k$ and $\tilde I_M - \tilde I_1 < N - M$.  $S$ and its inverse are in fact
the only two transformations satisfying these properties.  

\begin{SCfigure}
\includegraphics[width=6cm]{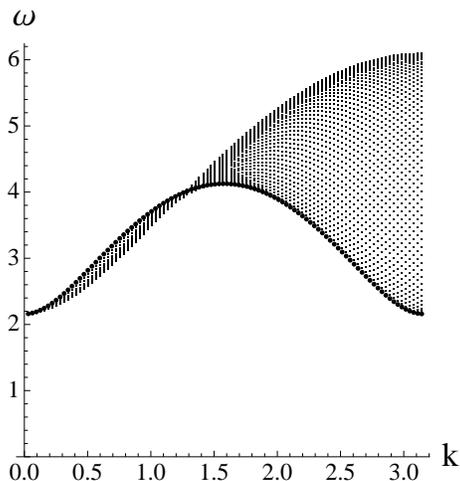}
\caption{The set of two-spinon excitations over the ground state of the $\Delta = 4$ chain, for $N = 200$ sites.  
Each point represents a state in the set of excitations defined as $A_0$ in the text.  
The set $A_1$ would lead to a very similar picture, shifted
by $\pi$ in momentum.  The darker points correspond to states in the set $A_0 \cap A_1$, for which we must explicitly 
prevent double-counting when summing for the structure factor.}
\label{fig:Spectrum_A0}
\end{SCfigure}

Consider now the set $A_i = \{ \{ I_j \} \bigm| |I_j - i| \leq \frac{N - M - 1}{2} \}$, {\it i.e.} the set of width $N - M$ of possible
quantum numbers, right-shifted by $i$.  For clarity, $A_0$ represents the usual set of quantum numbers used in the $XXX$ or gapless $XXZ$ cases
to get finite, real rapidities.
Consider now an element $x \in A_i \cup A_{i+1}$.  Since $\tilde I_1 - I_1 \geq 2$ and $\tilde I_M > \frac{N-M-1}{2} + i$
under the transformation $S$ defined above, 
we have $S(x) \notin A_i \cup A_{i+1} ~ \forall x$.  Taking $y \in (A_i \cup A_{i+1})/(A_{i+2} \cup A_{i+3})$, we have
$S(y) \in A_{i+2} \cup A_{i+3} ~\forall y$.  Therefore, as far as wavefunctions are concerned, we have
the equivalences $A_i \cup A_{i+1} \sim A_{i+2} \cup A_{i+3}$.  We can thus obtain a single enumeration of all wavefunctions
with purely real rapidities by restricting to $A_0$ and $A_1$, paying attention to double counting.  
The number of different wavefunctions is
thus given by the number of elements in $A_0$ and $A_1$, minus the number in the overlap $A_0 \cap A_1$, namely
\begin{equation}
2 \left( \begin{array}{c} N-M \\ M \end{array} \right) - \left( \begin{array}{c} N - M - 1 \\ M \end{array} \right)
= \frac{N}{N-M} \left( \begin{array}{c} N-M \\ M \end{array} \right).
\end{equation}
For the transverse structure factor at zero field, we are interested in states in the $M = N/2 - 1$ subsector.  
At finite size, the true ground state of the system has zero momentum, and the $N (N+2)/8$ states in $A_0$ are 
interpreted as the two-spinon states constructed above this ground state (see Figure \ref{fig:Spectrum_A0}).  
We can interpret the $N (N+2)/8$ states in $A_1$ as the set of two-spinon states built on the $\pi$ momentum quasi-degenerate 
ground state.  These would form a figure very similar to that obtained for $A_0$, but shifted by $\pi$.  
The overlap of the $A_0$ and $A_1$ sets is simply given by the set of states living on the darker line in Figure \ref{fig:Spectrum_A0}.
Counting states only once over this overlap region, the total number of states with real rapidities is 
thus in this case $\frac{N^2}{4}$ for all momenta considered together, and $\frac{N}{4}$ for each individual allowed momentum.
These are the two-spinon states of the finite chain, from which we will obtain the finite lattice structure factor later on.

\subsection{The infinite chain:  algebraic analysis}
We now turn to a complementary approach, which is valid strictly and only for the
infinite chain in zero magnetic field.  Known as the algebraic analysis, it provides
a construction of states and form factors from purely algebraic considerations, by exploiting the quantum affine symmetry 
of the model.  It is described in detail in \cite{JimboBOOK}, which contains essentially all the results we need.

As compared to the finite chain, the infinite XXZ antiferromagnet in the massive regime 
has a number of important characteristics.  First of all, the ground state becomes
exactly degenerate with the $\pi$-momentum Umklapped ground state.  Within the algebraic analysis, this translates into the existence
of two vacua $|\mbox{vac}\rangle_{(i)}$ whose label $i = 0,1$ differentiates two different possible 
asymptotic conditions, specifying the $z$-component of the spin on a chosen reference site.  Secondly, a separable subspace of the Hilbert space
can be identified, which is spanned by multispinon states \cite{FaddeevPLA85}.  The lore is that
all physical properties of the infinite chain can be derived using only the subspace 
${\cal F}$ of (up to denumerably infinite) spinon excitations over the two vacua.  This is in fact a remarkable statement, considering
that the true Hilbert space is only isomorphic to ${\cal F}$ in the subspace of denumerably finite spinon numbers.  This subspace
is spanned by states $\ket{\xi_m,\ldots,\xi_1}_{\epsilon_m,\ldots,\epsilon_1;(i)}$ with $j=1,\ldots,m$ 
spinons, each characterised by a pair $(\xi_j,\epsilon_j)$ with spectral parameter $\xi_j\in\{\mathbb{C} :|\xi_j|=1\}$ 
living on the complex unit circle and index $\epsilon_j\in\{-,+\}$ giving the spinon's spin orientation. 
Spinons are always created in (multiple) pairs by local spin flips, and two-spinon states are always
four-fold degenerate because of the different spinon spin orientation choices.

If we let $H_{\infty}$ denote the Hamiltonian of the infinite chain 
\footnote{Unfortunately, there exist different conventions for the Hamiltonian in the literature.  In the
algebraic approach, the usual choice is to use a ferromagnetic exchanged Hamiltonian, {\it e.g.} 
$H_{\infty} = -\frac{J}{4} \sum_j (\sigma^x_j \sigma^x_{j+1} 
+ \sigma^y_j \sigma^y_{j+1} + \Delta_{\infty} \sigma^z_j \sigma^z_{j+1})$, which we follow for convenience when
using this approach.  To make contact with the more typical physically relevant antiferromagnetic form (\ref{XXZ}), 
one should simply put $\Delta_{\infty} = -\Delta$, 
and rotate spins by $\pi$ around the $z$ axis on alternate sites, which shifts the total momentum by $\pi$
in the resulting structure factor.} 
and $T$ denote the translation operator by one site, we have that
\begin{eqnarray}
H_{\infty}\ket{\xi_m,\ldots,\xi_1}_{\epsilon_m,\ldots,\epsilon_1;(i)}&=E_{m}(\{\xi\}) \ket{\xi_m,\ldots,\xi_1}_{\epsilon_m,\ldots,\epsilon_1;(i)}\\
T \ket{\xi_m,\ldots,\xi_1}_{\epsilon_m,\ldots,\epsilon_1;(i)}&=e^{iP_{m}(\{\xi\})}\ket{\xi_m,\ldots,\xi_1}_{\epsilon_m,\ldots,\epsilon_1;(1-i)}
\end{eqnarray}
with 
\begin{equation}
E_{m}(\{\xi\})=\sum_{j=1}^{m}e(\xi_j)\,,\quad P_{m}(\{\xi\})=\sum_{j=1}^{m}p (\xi_j).
\end{equation}
$e(\xi_j)$ and $p(\xi_j)$ are respectively the energy and momentum of one spinon, with values
\begin{equation}
e^{-i p(\xi)}=\frac{1}{\xi}\frac{\Theta_{q^4}(q\xi^2)}{\Theta_{q^4}(q\xi^{-2})}\,,\quad e(\xi)=J\frac{1-q^2}{4q}\xi\frac{d}{d\xi}\log \tau(\xi)
\end{equation}
where 
\begin{eqnarray}
\Theta_{q}(w)&\equiv (w;q)_{\infty}(qw^{-1};q)_{\infty}(q;q)_{\infty}\\
(w;q)_{\infty}&\equiv\prod_{n=0}^{\infty}\left(1-wq^{n}\right).
\end{eqnarray}
Here, $q$ is the so-called deformation parameter of the quantum group
$U_q (sl_2)$, related to the anisotropic parameter by the formula $\Delta = - \Delta_{\infty}=-(q+q^{-1})/2$ ($-1<q<0$). 
Note that the translation operator $T$ maps the $m-$spinon states from one vacuum to the other:
it is therefore more convenient to work with translational invariant spinon states 
\begin{equation}
\fl
\ket{\xi_m,\ldots,\xi_1;p}_{\epsilon_m,\ldots,\epsilon_1}=\frac{1}{\sqrt{2}}\left[\ket{\xi_m,\ldots,\xi_1}_{\epsilon_m,\ldots,\epsilon_1;(0)}
+e^{ip}\ket{\xi_m,\ldots,\xi_1}_{\epsilon_m,\ldots,\epsilon_1;(1)}\right]\,,
\end{equation}
with $p=0,\pi$.  In particular, the two translational invariant vacua, {\it i.e.} no spinons, read
\begin{eqnarray}
\ket{0}&=\frac{1}{\sqrt{2}}\left[\ket{\mbox{vac}}_{(0)}+\ket{\mbox{vac}}_{(1)}\right]\\
\ket{\pi}&=\frac{1}{\sqrt{2}}\left[\ket{\mbox{vac}}_{(0)}-\ket{\mbox{vac}}_{(1)}\right],
\end{eqnarray}
and correspond to the infinite size limit of the two quasi-degenerate ground states of the finite lattice.
Translational invariant states are eigenstates of the translation operator $T$,
\begin{equation}
T \ket{\xi_m,\ldots,\xi_1;p}_{\epsilon_m,\ldots,\epsilon_1}
=e^{i(P_{m}(\{\xi\})+p)}\ket{\xi_m,\ldots,\xi_1;p}_{\epsilon_m,\ldots,\epsilon_1}.
\end{equation}
Finally, a resolution of the identity within subspace ${\cal F}$ can be written in terms of translational invariant spinons as
\begin{equation}
\fl
\mathbb{I} =\sum_{m\geq 0}\sum_{\epsilon_1,\ldots,\epsilon_m}\frac{1}{m!}\sum_{p=0,\pi}\oint\prod_{i=1}^m \frac{d\xi^{2}_i}{2\pi i\xi^{2}_i }
\ket{\xi_m,\ldots,\xi_1;p}_{\epsilon_m,\ldots,\epsilon_1}\,_{\epsilon_1,\ldots,\epsilon_m}\bra{\xi_1,\ldots,\xi_m;p}
\label{eq:resol}
\end{equation}
where the contour integral is such that the square of the spectral parameter $\xi^2_i$ {\it covers the unit circle once}. 
Note that the measure in the resolution of the identity is different from the one given in \cite{JimboBOOK} and normally 
used in the literature (see for instance \cite{BougourziPRB57,KarbachPRB55}). The reason for this is to keep the momentum of the spinon 
within its physical range of width $\pi$ \cite{FaddeevPLA85}  and to obtain the usual spinon dispersion relation (see for
example \cite{TakahashiBOOK}), in accordance with the physical picture given by Bethe Ansatz on the finite lattice.
The different resolution of the identity stems from the fact that, within the algebraic approach, states with spectral
parameters $\xi$ and $-\xi$ should be identified (see equation (A.12) in \cite{JimboBOOK}).

It is mathematically more convenient to work with the following parametrisation for the spectral parameter $\xi$:
\begin{equation}
\xi=ie^{i\pi\beta/2K}\,,\quad -K\leq \beta< K,
\end{equation}
with $\beta$ defined modulo $4K$, 
with $K\equiv K(k)$ the complete elliptic integral of the first kind with elliptic modulus $k$ (not to be confused with the external momentum,
when we deal with the finite lattice). 
The deformation parameter $q$ will be minus the elliptic nome $q_{e}$
\begin{equation}
-q=q_{e}=e^{-\frac{\pi K'}{K}}
\end{equation}
with $K'\equiv K(k')$ and $k'\equiv\sqrt{1-k^2}$ the complementary elliptic modulus. The spinon energy and momentum then read
\begin{equation}
e(\beta)=I\mbox{dn}(\beta), \quad
p(\beta)=\mbox{am}(\beta)+\frac{\pi}{2}\,,\quad I\equiv \frac{J K}{\pi}\sinh\left(\frac{\pi K'}{K}\right)
\end{equation}
where $\mbox{dn}(x)$ and $\mbox{am}(x)$ are the Jacobi elliptic functions with elliptic modulus $k$. 
The spinon dispersion relation is then explicitly given by
\begin{equation}
e_1(p)=I\sqrt{1-k^2\cos^2(p)}\,,\quad 0\leq p\leq \pi,
\end{equation}
where the spinon momentum $p$ take values in the domain $p\in [0,\pi]$.  This dispersion relation is illustrated by the
darker line in Figure \ref{fig:Spectrum_A0} (up to a shift of $Ik'$, since the second spinon is then sitting on zero momentum), 
and becomes the well-known sine curve in the isotropic limit.

As it is pointed out in \cite{JimboBOOK}, these formulas correspond to the ones appearing for example in \cite{JohnsonPRA8,BabelonNPB220}, 
whereas one might have thought that the range of the spinon momentum was $2\pi$ in the former approach.  
Although mathematically equivalent when the states are correctly identified and the dynamical constraints
correctly solved, we prefer to work 
with the spinon momentum $p(\beta)$ confined within the interval $[0,\pi]$ to make contact with the finite $N$ Bethe Ansatz clearer. 
This explains our choice $-K\leq \beta\leq K$ and the change of measure in the resolution of the identity (\ref{eq:resol}).

In the thermodynamic limit, the intermediate states we will use will be constructed by creating a pair of spinons
over either of the translationally invariant vacua.  On the finite lattice, these states then correspond to the
two-spinon states created on the two quasi-degenerate ground states.  Formally, there then exists a double counting
of states on a line (darker line in Figure \ref{fig:Spectrum_A0}) in the algebraic approach, but this set of relative measure zero
does not influence the result for the TDSF.

Let us now turn to the construction of the contribution to the TDSF coming from the two-spinon intermediate states. 
As before, we treat the finite lattice and infinite one in turn.

\section{The transverse dynamical structure factor}
\label{sec:TDSF}
The gapped Heisenberg chain, as detailed above, has two quasi-degenerate ground states at finite $N$ which
become exactly degenerate in the thermodynamic limit $N \rightarrow \infty$.  The chain thus has a tendency
to spontaneously develop a staggered magnetization along the $z$ axis, selecting for example one of the
states $|vac\rangle_{(i)}$ with broken translational symmetry.  We here wish to begin by discussing the effects of this
tendency to spontaneously order on the TDSF.

When spontaneous order develops, due to the broken translational invariance of the ground state, 
the period of the magnetic structure is doubled, and the magnetic
Brillouin zone is therefore halved.  States and excitations can thus be written as modes over this reduced Brillouin zone
(since the Brillouin zone is halved, the number of modes per momentum is doubled, {\it i.e.} we could separately classify
sites into odd and even ones).
One might expect the TDSF to share this $k \rightarrow k + \pi$ periodicity, but this is not the case.  Namely,
consider calculating a structure factor on either of the two possible ordered states $|0\rangle$ or $|1\rangle$
(which can be here taken to represent the finite-size equivalents of $|\mbox{vac}\rangle_{(i)}$):
\begin{eqnarray}
S^{ab}_i (k, \omega) = \frac{1}{N} \sum_{j,j'} e^{-i k(j-j')} \int_{-\infty}^{\infty} dt e^{i\omega t} 
\langle i | S^a_j (t) S^b_{j'}(0) | i \rangle.
\end{eqnarray}
%By using the property $\langle i | S^a_{j+2} (t) S^b_{j'+2}(0) | i \rangle = \langle i | S^a_j (t) S^b_{j'}(0) | i \rangle$, 
We can separate this into the following contributions:
\begin{eqnarray}
S^{ab}_i (k,\omega) = F^{ab;i}_{e} (k,\omega) + e^{-ik} F^{ab;i}_{o} (k,\omega)
\label{Sabi}
\end{eqnarray}
with $F^{ab;i}_{e/o} (k,\omega) = F^{ab;i}_{e/o,0} (k,\omega) + F^{ab;i}_{e/o,1} (k,\omega)$ being the contributions from
correlators involving pairs of sites at an even/odd distance, which are further separated in the functions
\begin{eqnarray}
F^{ab;i}_{e,\epsilon} (k,\omega) = \sum_n e^{-ik 2n} \int_{-\infty}^{\infty} dt e^{i\omega t} \langle i | S^a_{2n + \epsilon} (t)
S^b_{\epsilon} (0) | i\rangle, \nonumber \\
F^{ab;i}_{o,\epsilon} (k,\omega) = \sum_n e^{-ik 2n} \int_{-\infty}^{\infty} dt e^{i\omega t} \langle i | S^a_{2n + 1 + \epsilon} (t)
S^b_{\epsilon} (0) | i\rangle, 
\end{eqnarray}
with $\epsilon = 0,1$ labelling the sublattice of the base site.  
All functions $F^{ab;i}$ are then manifestly $\pi$-periodic in momentum,
$F^{ab;i}_{e/o,\epsilon} (k+\pi, \omega) = F^{ab;i}_{e/o,\epsilon} (k, \omega)$, but $S^{ab}_i (k, \omega)$ does not
share that property since the $e^{-ik}$ term in (\ref{Sabi}) does not vanish.  The points $k$ and $k + \pi$
therefore remain inequivalent in the structure factor.  The conclusion remains the same for the combination
$S^{ab} = \frac{1}{2} (S^{ab}_0 + S^{ab}_1)$ which we consider at infinite size
(which thus corresponds to the TDSF at a temperature much higher than the splitting of the two ground states
(which is zero), but much lower than the gap);  in fact, the structure factor is the same on
either ordered state, so all linear combinations give the same result (at finite size, we calculate the
true zero-temperature TDSF, namely on the true ground state only).  The limits $N \rightarrow \infty$ and
$T \rightarrow 0$ commute, and spontaneous symmetry breaking does not affect the periodicity in momentum
of the structure factor, which remains $2\pi$.  This is similar to what happens in the presence of a staggered field, see
{\it e.g.} the discussion in \cite{WangPRB66}.  With this in mind, let us now explicitly construct the TDSF,
first on the lattice, then in the continuum.

\subsection{Finite chain}
On a finite lattice, the transverse dynamical structure factor (\ref{TDSF}) is defined for discrete values of 
momenta $k = 2\pi n /N$ with $n \in \mathbb{Z}$.  By periodicity over the full length of the chain, we can restrict the momentum
to the first Brillouin zone $n \in \{ 0, N-1 \}$.  Using the Fourier transform of the spin operators
$S^a_k = \frac{1}{\sqrt{N}} \sum_j e^{-ikj} S^a_j$ and introducing a formal sum over intermediate states
$|\alpha \rangle$, the TDSF can be written in a Lehmann representation as
\begin{eqnarray}
S^{-+}(k, \omega) = 2\pi \sum_{\alpha} | \langle 0 | S^-_k | \alpha \rangle|^2 
\delta (\omega - E_{\alpha} + E_0)
\label{TDSF_Lehmann}
\end{eqnarray}
where $E_0$ is the energy of the ground state $|0\rangle$, and $E_{\alpha}$ is the energy of
state $|\alpha\rangle$.  The form factors are then computed using the known determinant representations
\cite{KitanineNPB554,KitanineNPB567} once the Bethe equations have provided the rapidities of both
eigenstates involved (see {\it e.g.} \cite{CauxPRL95,CauxJSTATP09003} for a more extensive discussion).

\begin{SCfigure}
\includegraphics[width=7cm]{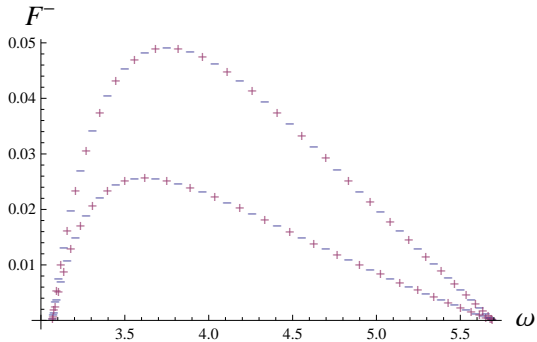}%ffmin-N120D4low.eps}
\caption{Square form factors coming from two spinon excitations over the true zero momentum ground
state ($-$) and over the $\pi$-momentum quasi-degenerate ground state ($+$), for momentum $k = \pi/4$
(lower curves) and $k = 3\pi/4$ (upper curves), for $\Delta = 4$.  }
%The structure factor, which is the same
%when computed on either ground state, is asymmetric with respect to the $\pi/2$ momentum line.}
\label{A0A1equalSF}
\end{SCfigure}

The full set of intermediate states 
$\{ | \alpha \rangle \}$ for a finite lattice separates into many different classes of states,
only some of which are important for the TDSF.  In the case at hand, the full Lehmann sum in (\ref{TDSF_Lehmann}) can
be truncated very efficiently by considering only states with real solutions to the Bethe equations,
which correspond to the two-spinon states described earlier.  The structure factor can thus be
understood as sums of contributions coming from spinons built on the zero-momentum true ground state
(the set $A_0$ of excitations), and of those built on the $\pi$-momentum quasi-degenerate ground state
(the $A_1$ set).  These two contributions turn out to be equal to one another (up to $1/N$ corrections).
This is shown in Figure \ref{A0A1equalSF}, and illustrates two important things.  
First, both base states ($0$ and $\pi$ momentum ground states) in fact yield the same structure factor in the
continuum limit, so which linear combination of these states is taken is immaterial in the definition of
the structure factor.  Second, the structure factor will clearly be asymmetric with respect to the 
$\pi/2$ momentum line.
In the Ising limit, the form factors do not depend anymore on energy, and the structure factor thus precisely follows the density of
states.  Since the two sets of form factors from two-spinon states are then identical (up to $1/N$ corrections) 
modulo a shift of $\pi$ in momentum, only then does the TDSF becomes symmetric with respect to the $k = \pi/2$ line.  
On the other hand, in the isotropic limit, states in the set $A_1/ (A_0 \cap A_1)$ contain
one rapidity equal to $\pi/2$ (or $\infty$ in the usual parametrization at the isotropic point),
which represents the action of the total spin lowering operator $S^-_{k = 0}$ at zero momentum.  Since the Bethe
wavefunctions are highest-weight states of the $SU(2)$ total spin rotation symmetry, all the contributions
from those states vanish due to the selection rules, leaving only the $N(N+2)/8$ states in the set
$A_0$ to provide nonzero contributions.  The TDSF then recovers its well-known `Viking helmet' shape,
vanishing at $k = 0$ and peaking at $k = \pi$, and thus having maximal anisotropy with respect to the $\pi/2$ momentum line.

Our results for the lattice TDSF are explicitly plotted together with those in the thermodynamic
limit in Figure \ref{DSF2}.  We postpone discussion of these until later, but now look at the
calculation of the TDSF of the infinite chain.

\subsection{Infinite chain}
\begin{figure}
\begin{tabular}{cc}
\psfrag{Rminus}{{\large $\mathcal{R}_{-}$}}
\psfrag{Rplus}{{\large $\mathcal{R}_{+}$}}
\psfrag{Q}{$Q$}
\psfrag{w}{$\omega/I$}
\includegraphics[height=5cm, width=6cm ]{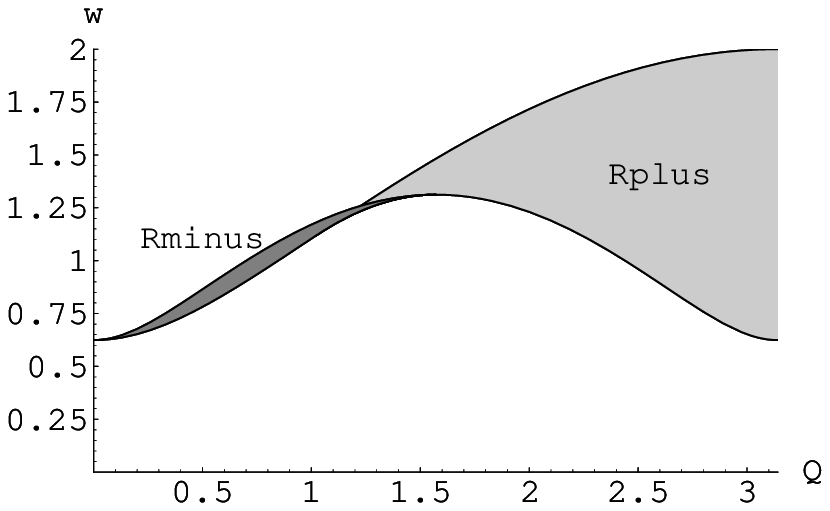}
&
\psfrag{Cplus}{{\large $\mathcal{C}_{+}$}}
\psfrag{Q}{$Q$}
\psfrag{w}{$\omega/I$}
\includegraphics[height=5cm, width=6cm ]{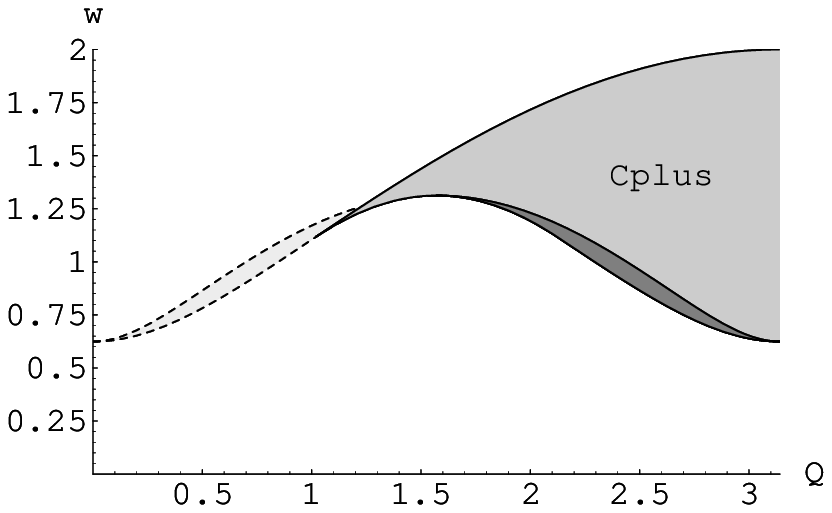}
\end{tabular}
\caption{Left: the two-spinon continuum $\mathcal{R}$ for $k=0.95$ ($\Delta= 3.601$).
Right:  the $\mathcal{C}_{+}$ continuum for the same anisotropy.}
\label{fig:regions}
\end{figure}

In the continuum, the two-spinon part of the TDSF has already been intensively investigated 
\cite{BougourziPRB57}.  For clarity and completeness, we here reproduce the outline of the derivation,
which involves some slight differences with the earlier treatment.  We start by writing the TDSF as 
\begin{equation}
S^{-+} (Q, \omega) = \sum_{j=-\infty}^{\infty} e^{-iQj} \int_{-\infty}^{\infty} dt e^{i\omega t} \langle S^-_{j+1}(t) S^+_1(0) \rangle
\end{equation}
where $\bracket{\cdots}=(1/2)\sum_{p=0,\pi}\bra{p}\cdots\ket{p}$. 
To decompose this into the different spinon contributions, one simply inserts the resolution 
of the identity (\ref{eq:resol}) between the spin operators. 
Upon defining the form factors  
$X^{(i)}_{\{\epsilon\}}(\{\xi\})=\,_{(i)}\bra{\mbox{vac}}\sigma^{+}_{1} \ket{\xi_m,\epsilon_m;\cdots ;\xi_1,\epsilon_1}_{(i)}$, 
the TDSF takes the form
\begin{equation}
S^{-+}(Q,\omega)=\sum_{m \mbox{ even }\geq 0}S^{-+}_{(m)}(Q,\omega)
\end{equation}
with
\begin{eqnarray}
\fl
S^{-+}_{(m)}(Q,\omega)&= \frac{\pi^2}{ m!}\oint\prod_{i=1}^m \frac{d\xi^2_i}{2\pi i\xi^2_i }\delta[\omega-E_m(\{\xi\})]\sum_{p=0,\pi}\delta_{(2\pi)}[Q+p-P_m(\{\xi\})] B^{(\sigma)}(\{\xi\}) \nonumber \\
\fl
B^{(\sigma)}(\{\xi\})&\equiv\sum_{\{\epsilon\}}B^{(\sigma)}_{\{\epsilon\}}(\{\xi\})\,,\quad B^{(\sigma)}_{\{\epsilon\}}(\{\xi\})\equiv\left| X^{(0)}_{\{\epsilon\}}(\{\xi\})-\sigma X^{(1)}_{\{\epsilon\}}(\{\xi\})\right|^2\,,
\label{eq:Smp_and_B}
\end{eqnarray}
with $\sigma=-e^{ip}$.  We refer to $S^{-+}_{(m)}(Q,\omega)$ as the $m-$spinon TDSF, or simply  TDSF$_{(m)}$, 
which is simply the contribution to the TDSF coming from $m$-spinon intermediate states.
Using the $\beta$ parametrisation, the TDSF$_{(m)}$ becomes
\begin{eqnarray}
\fl
\hspace{1cm}S^{-+}_{(m)}(Q,\omega)=\frac{\pi^2}{m!}\left(\frac{1}{2K}\right)^{m}\int_{-K}^{K}\int_{-K}^{K} d\beta_1\cdots d\beta_m\,
\delta[\omega-E_m(\{\beta\})] \times \hspace{1cm}\nonumber \\
\hspace{1cm}\times \sum_{p=0,\pi}\delta_{(2\pi)}[Q+p-P_m(\{\beta\})] B^{(\sigma)}(\{\beta\}).
\label{eq:mDSF_beta}
\end{eqnarray}
Let us now concentrate on two-spinon states.  The energy of such states is simply given by the sum of 
the energies of the two spinons, 
\begin{equation}
e_2(p_1,p_2)=I\sqrt{1-k^2\cos^2(p_1)}+I\sqrt{1-k^2\cos^2(p_2)},
\end{equation}
whereas the total momentum of the two-spinon state is simply $Q=p_1+p_2\in[0,2\pi]$. 
One can write the two-spinon energy as a function of $Q$ and a parameter $\lambda=\frac{1}{2}(p_1-p_2)$ with 
$\lambda \in [-\mbox{Min}(Q/2, \pi - Q/2), \mbox{Min}(Q/2, \pi - Q/2)]$. 
The two-spinon dispersion region forms a continuous region $\mathcal{R}=\mathcal{R}_{-}\cup\mathcal{R}_{+}$ 
as in the left panel of Fig. \ref{fig:regions}, which is characterized by a very narrow band near $Q = 0$ and a broad
continuum around $Q = \pi$.  This is the usual two-spinon continuum as it appears in the context 
of Bethe Ansatz (see for instance \cite{TakahashiBOOK}, and recall the discussion in the previous section).  
Some detailed comments are however worthwhile to make.  
If we define $\kappa=(1-k')/(1+k')$ and
\begin{equation}
\omega_{\pm}(Q)\equiv \frac{2 I}{1+\kappa}\sqrt{1+\kappa^2\pm2\kappa\cos(Q)}, \quad
\omega_0(Q)\equiv \frac{2I}{1+\kappa}\sin(Q),
\end{equation}
the lower boundary of region $\mathcal{R}$ for $Q \in [0, \pi]$ is given by
\begin{eqnarray}
\omega_{lo} (Q) = \left\{ \begin{array}{cc}
\omega_- (Q), & Q \in [0, Q_{\kappa}], \\
\omega_0 (Q), & Q \in [Q_{\kappa}, \pi/2], \\
\omega_+ (Q), & Q \in [\pi/2, \pi] \end{array} \right.
\end{eqnarray}
where $Q_{\kappa} = \mbox{acos} (\kappa)$.  In other words, between $Q = 0$ and $Q = Q_{\kappa}$, the lower boundary
is defined by setting $\lambda = 0$ so $p_1 = p_2 = Q/2$ (here and in what follows, we can always interchange
$p_1$ and $p_2$, and the solution to the dynamical constraints will always have this degeneracy).  
Between $Q = Q_{\kappa}$ and $Q = \pi/2$, it is however obtained by setting
$\lambda = \frac{1}{2} \mbox{acos} \frac{Q}{\kappa}$.  Finally, between $Q = \pi/2$ and $\pi$, 
it is obtained by setting $\lambda = Q/2$ so $p_1 = Q, p_2 = 0$, so we here simply fall back on the
spinon dispersion relation (shifted by $Ik'$ since the second spinon sits at zero momentum). 

The upper boundary of $\mathcal{R}$ for $Q \in [0, \pi]$ is given by
\begin{eqnarray}
\omega_{up} (Q) = \left\{ \begin{array}{cc}
I k' + I\sqrt{1 - k^2 \cos^2 Q}, & Q \in [0, Q_c], \\
\omega_- (Q), & Q \in [Q_c, \pi] \end{array} \right.
\end{eqnarray}
where $Q_c$ is obtained from solving the quartic equation
\begin{eqnarray}
\fl
\cos^4 Q_c - 4 \cos^3 Q_c + 2 (4/k^2 - 1) \cos^2 Q_c - 4 (2/k^2 - 1) \cos Q_c + 1 = 0,
\end{eqnarray} 
whose solution can easily be found in closed form,
\begin{eqnarray}
\cos Q_c = 1 - 4 (2/3)^{1/3} \frac{(1-k^2)^{2/3}}{k \Sigma} + (2/3)^{2/3} \frac{(1-k^2)^{1/3}}{k} \Sigma,
\end{eqnarray}
with $\Sigma = (\sqrt{3(32 - 5k^2)} - 9k)^{1/3}$.
Thus, the upper boundary is simply defined by setting $\lambda = Q/2$ between $Q = 0$ and $Q = Q_c$, 
meaning that $p_1 = Q$ and $p_2 = 0$.  Afterwards $\lambda$ becomes 0, so the two spinons share the
same momentum.  The region $Q \in [\pi, 2\pi]$ is simply described by taking $p_i \rightarrow \pi - p_i$.

Within region $\mathcal{R}_+$, the two-spinon states are therefore
ordered in increasing energy with decreasing $|\lambda|$.  In
region $\mathcal{R}_-$, this ordering is reversed;  this region exists due to the change of sign of the second
momentum derivative of the spinon dispersion relation, which is greater than zero for small $k$ but becomes negative
at an anisotropy-dependent value $k_c = \mbox{acos}\frac{1}{\sqrt{1 + k'}}$.  
Convolving two spinon dispersion relations thus produces this `folding' region where $\mathcal{R}_-$ and $\mathcal{R}_+$
overlap.  This region simply disappears at the isotropic point, since the
spinon dispersion relation then has strictly negative curvature.  However, a similar region exists for a generic
XXZ chain in a field (see the discussion in \cite{PereiraJSTATP08022}).  

For computational purposes and to make contact with earlier results, it is convenient to observe that
if the region $\mathcal{R}_{-}$ is reflected around $Q=\pi/2$, we obtain the continuous region depicted 
in the right panel of Fig. \ref{fig:regions}, which corresponds to the sheet $~\mathcal{C}_{+}$ in \cite{BougourziPRB57}.
Note that this reflection is such that region $\mathcal{R}_-$ is never overlapping with $\mathcal{R}_+$, but
fits precisely under it.  The resulting sheet is such that each point refers to a single eigenstate
(up to the trivial symmetry $p_1 \leftrightarrow p_2$), in other
words solutions to dynamical constraint equations are unique within the physical region $\beta \in [-K, K]$ 
up to simple permutation $\beta_1 \leftrightarrow \beta_2$.  
This was not the case before, since $\mathcal{R}_-$
and $\mathcal{R}_+$ had a nonzero overlap.

%\begin{figure}[h]
%\psfrag{Rminus}{{\large $\mathcal{R}_{-}$}}
%\psfrag{Rplus}{{\large $\mathcal{R}_{+}$}}
%\psfrag{Q}{$Q$}
%\psfrag{w}{$\omega/I$}
% \includegraphics[height=5cm, width=7cm ]{regionR_k0.95.eps}
%\caption{Two spinon continuum $\mathcal{R}$ for $k=0.95$ ($\Delta=-3.601$)}
%\label{fig:regionR}
%\end{figure}
%\begin{figure}[h]
%\psfrag{Cplus}{{\large $\mathcal{C}_{+}$}}
%\psfrag{Q}{$Q$}
%\psfrag{w}{$\omega/I$}
% \includegraphics[height=5cm, width=7cm ]{regionC_k0.95_v2.eps}
%\caption{Two spinon continuum $\mathcal{C}_{+}$ for $k=0.95$ ($\Delta=-3.601$)}
%\label{fig:regionC}
%\end{figure}
The boundaries of the sheet $\mathcal{C}_{+}$ in the interval $Q\in[0,\pi]$ are given by
\begin{equation}
\Omega_{\mbox{up}}(Q)=\omega_{-}(Q)
\end{equation}
for the upper boundary, and
\begin{equation}
\Omega_{\mbox{lo}}(Q)=\left\{
\begin{array}{cc}
\omega_0(Q)\,,  & Q\in[Q_{\kappa},\pi-Q_{\kappa}]\\
\omega_{+}(Q)\,, &  Q\in[\pi-Q_{\kappa},\pi]
\end{array}\right.
\end{equation}
for the lower boundary.  As in \cite{BougourziPRB57}, 
we will calculate the TDSF using the sheet $\mathcal{C}_{+}$ but, as we will explain below, 
one must be careful when dealing with the energy-momentum relations. For later use, we also 
introduce the sheet $\mathcal{C}_{-}$ as the reflection of the sheet $\mathcal{C}_{+}$ around $Q=\pi/2$.

To calculate the two-spinon TDSF, we will need to explicitly solve the two-spinon energy and momentum relations
\begin{equation}
Q=\mbox{am}(\beta_{1})+\mbox{am}(\beta_{2})+\pi, \quad
\frac{\omega}{I}=\left[\mbox{dn}(\beta_{1})+\mbox{dn}(\beta_{2})\right].
\label{eq:ek}
\end{equation}
We follow the derivation in \cite{BougourziPRB57} using addition formulas of elliptic functions. 
Upon introducing the new variables $\beta_{\pm}=(\beta_1\pm\beta_{2})/2$, the solution to the set of equations 
(\ref{eq:ek}) in the sheet $\mathcal{C_{+}}$ is given by 
\begin{eqnarray}
\beta^{(+)}_{+}(Q,\omega)&=-\frac{1+\kappa}{2}F\left[\mbox{arcsin}\left(\frac{\omega_0}{\omega}\right),\kappa\right], \nonumber \\
\beta^{(+)}_{-}(Q,\omega)&=\mbox{dn}^{-1}\left(\frac{1+\cos(Q)}{|\sin(Q)|}\sqrt{\frac{\omega^2-\kappa\omega_0^2+T}{\omega^2+\kappa\omega_0^2-T}},k\right)
\end{eqnarray}
with 
\begin{equation}
T=T(Q,\omega) \equiv \sqrt{\omega^2-\kappa^2\omega_0^2}\sqrt{\omega^2-\omega_0^2}.
\end{equation}
Within the interval $-K\leq \beta_1,\beta_2\leq K$ another solution is naturally obtained by the transformation 
$\beta_1\leftrightarrow\beta_2$, which keeps the set of  equations (\ref{eq:ek}) unchanged.

While we have obtained the solutions within the sheet $\mathcal{C}_+$, the two-spinon continuum is the region 
$\mathcal{R}=\mathcal{R}_{-}\cup\mathcal{R}_{+}$.  To find the corresponding solution within $\mathcal{R}$, we 
note from the set of eqs (\ref{eq:ek}) that solutions for $(Q,\omega)$ and $(\pi-Q,\omega)$ are related by
\begin{equation}
\beta_1(Q,\omega)=-\beta_1(\pi-Q,\omega), \quad
\beta_2(Q,\omega)=-\beta_2(\pi-Q,\omega)-2K.
\end{equation}
While the minus sign in these expressions is not important when evaluating the $B$ functions, 
the shift in $2K$ is important, since it changes the relative sign of the spectral parameters $\xi_1$ and $\xi_2$. Defining
\begin{eqnarray}
\beta^{(\sigma)}_{+}(Q,\omega)&=-\frac{1+\kappa}{2}F\left[\mbox{arcsin}\left(\frac{\omega_0}{\omega}\right),\kappa\right], \nonumber \\
\beta^{(\sigma)}_{-}(Q,\omega)&=\mbox{dn}^{-1}\left(\frac{1+\sigma\cos(Q)}{|\sin(Q)|}\sqrt{\frac{\omega^2-\kappa\omega_0^2+T}
{\omega^2+\kappa\omega_0^2-T}},k\right),
\label{solutions}
\end{eqnarray} 
the solutions in the two-spinon continuum $\mathcal{R}$ are given by
\begin{eqnarray}
(\beta_1,\beta_2)&=[\beta^{(+)}_1(Q,\omega),\beta_2^{(+)}(Q,\omega)]\,&\quad (Q,\omega)\in\mathcal{R}_{+}\\
(\beta_1,\beta_2)&=[\beta^{(-)}_1(Q,\omega),\beta_2^{(-)}(Q,\omega)-2K]\,&\quad(Q,\omega)\in\mathcal{R}_{-}.
\end{eqnarray}
We can now substitute these into the expressions for the form factors to obtain the TDSF.  
Recalling formula (\ref{eq:mDSF_beta}),  the projection of the TDSF onto the two-spinon band takes the form
\begin{eqnarray}
S^{-+}_{(2)}(Q,\omega)= \frac{1}{2}\left(\frac{\pi}{2K}\right)^{2} \int_{-K}^{K}\int_{-K}^{K} d\beta_1 d\beta_2\,
\delta[\omega-E_2(\{\beta\})] \times \nonumber \\
\times \sum_{p=0,\pi} \delta_{(2\pi)}[Q+p-P_2(\{\beta\})] B^{(\sigma)}(\{\beta\})\,.
\label{eq:DSF_twospinon}
\end{eqnarray}
The two-spinon form factors involved in (\ref{eq:DSF_twospinon}) appeared in the definition of the weights $B^{(\sigma)}(\{\beta\})$
in equation (\ref{eq:Smp_and_B}). 
Due to spin conservation, only the form factor with spin orientation $(\epsilon_1,\epsilon_2)=(+,+)$ gives a non-trivial contribution. 
This is related to the form factor with $(-,-)$ through $_{(i)}\langle \mbox{vac} | \sigma^{\pm} | \xi_2, \xi_1 \rangle_{--;(i)}
= _{(1-i)}\langle \mbox{vac} | \sigma^{\mp} | \xi_2, \xi_1 \rangle_{++;(1-i)}$.  All these form factors are obtained
from \cite{JimboBOOK}, and the fundamental building block reads
\begin{eqnarray}
\fl
X^{(i)}(\xi_2,\xi_1)\equiv\,_{(i)}\bra{\mbox{vac}}\sigma_1^{+}\ket{\xi_2,\xi_1}_{-\,-;(i)} \nonumber \\
\fl\quad=(-q)^{1-i}\xi^{1-i}_1\xi^{2-i}_2(q^2;q^4)_{\infty}(q^4;q^4)^{3}_{\infty}\rho^2\frac{\gamma(\xi^2_2/\xi^2_1)}
{\prod_{k=1}^2\Theta_{q^4}(\xi_k^{-2}q^3)}\Theta_{q^8}(-\xi_1^{-2}\xi_2^{-2}q^{4i})
\end{eqnarray}
with
\begin{eqnarray}
&(w;q,p)_{\infty}\equiv \prod_{n,m=0}^{\infty}\left(1-wq^{n}p^{m}\right)\,,\nonumber \\
&\gamma_{\sigma}(w)\equiv\frac{((-q)^{1+\sigma} q^{4}w;q^4,q^4)_{\infty}((-q)^{1+\sigma} w^{-1};q^4,q^4)_{\infty}}{((-q)^{3+\sigma} q^{4}w;q^4,q^4)_{\infty}((-q)^{3+\sigma} w^{-1};q^4,q^4)_{\infty}}
\end{eqnarray}
so that  $\gamma(w)\equiv\gamma_{-}(w)$ and $\rho^2\equiv \gamma_{+}(q^{-2})$. Using standard definitions and properties of 
Jacobi elliptic functions, one obtains the following expression for $B^{(\sigma)}(\beta_{\pm})$:
\begin{equation}
\fl
B^{(\sigma)}(\beta_{\pm})=\left(\frac{2K}{\pi}\right)^2\frac{\vartheta^2_{A}(\beta_{-})}{\vartheta_{d}^2(\beta_{-})}\frac{\mbox{dn}^2(\beta_{-})}{1- k^2\mbox{sn}^2(\beta_{-})\mbox{sn}^2(\beta_{+})}\left[k'\,\delta_{\sigma,-}+\mbox{dn}^2(\beta_{+})\delta_{\sigma,+}\right]
\end{equation}
where we have defined
\begin{equation}
\vartheta^2_{A}(\beta)\equiv \exp\left[-\sum_{k=1}^{\infty}\frac{e^{ k \epsilon}}{k}\frac{\cosh(2k\epsilon)\cos(2\beta k\epsilon/K')-1}{\sinh(2k \epsilon)\cosh(k\epsilon)}\right]
\end{equation}
with $\epsilon=\frac{\pi K'}{K}$, and where $\vartheta_{d}(\beta_{-})$ is the Neville theta function. 
Note that $B^{(\sigma)}(\beta_{\pm})$ is invariant under the transformation $\beta_{1}\leftrightarrow \beta_{2}$.

From  (\ref{eq:DSF_twospinon}), we can see that the TDSF$_{(2)}$ consists of the sum of the two-spinon region $\mathcal{R}$ 
weighted by $B^{(-)}(\{\beta\})$ plus the same region shifted by $\pi$ and weighted by $B^{(+)}(\{\beta\})$. 
By noticing that a shift in $2K$ in one of the parameters $\beta$ implies a change in the weights $B^{(-)}\leftrightarrow B^{(+)}$, 
the preceding description is equivalent of weighting the sheets $C_{\sigma}$ with the weights $B^{(\overline{\sigma})}$ for 
$\sigma=\pm$ and using as the solution to the energy-momentum equations the expressions (\ref{solutions}) {\it without} 
a shift of $2K$. Evaluation of $B^{(\overline{\sigma})}(\beta^{(\sigma)}_{\pm}(Q,\omega))\equiv B_{\sigma}^{(\overline{\sigma})}(Q,\omega)$ 
using such expressions gives
\begin{eqnarray}
\fl
B_{\sigma}^{(\overline{\sigma})}(Q,\omega)=\left(\frac{2K(\kappa)}{\pi}\right)^2\frac{1+\sigma\cos(Q)}{\omega_0^2}\frac{\vartheta_{A}^2[\beta^{(\sigma)}_{-}(Q,\omega)]}{\vartheta_{d}^2[\beta^{(\sigma)}_{-}(Q,\omega)]}\times \nonumber \\ 
\times \left[\frac{1-\kappa}{1+\kappa}[\omega^2+\kappa\omega_0^2+T]\delta_{\overline{\sigma},-}+[\omega^2-\kappa\omega_0^2+T]\delta_{\overline{\sigma},+}\right].
\end{eqnarray}
Upon defining
\begin{equation}
\fl
J_{\sigma}(Q,\omega)=\left(\frac{2K}{\pi}\right)^{2}\left|\frac{\partial E}{\partial \beta_{1}}\frac{\partial P}{\partial \beta_{2}}
-\frac{\partial E}{\partial \beta_{2}}\frac{\partial P}{\partial \beta_{1}}\right|_{\beta_{\pm}=\beta^{(\sigma)}_{\pm}(Q,\omega)}
=2\left(\frac{ 2K(\kappa)}{\pi}\right)^2 \frac{\omega T W_{\sigma}}{\omega_0^2}
\end{equation}
with $\omega_0\equiv \omega_0(Q)$, $T=T(Q,\omega)$ and $W_{\sigma}=W_{\sigma}(Q,\omega)$ with
\begin{eqnarray}
W_{\sigma}=W_{\sigma}(Q,\omega)&\equiv \sqrt{\kappa^2\frac{\omega_0^4}{\omega^4}-\left(\frac{T}{\omega^2}+\sigma\cos(Q)\right)^2}
\end{eqnarray}
and considering the multiplicity of the solutions, we can finally write
\begin{equation}
S^{-+}_{(2)}(Q,\omega)=\sum_{\sigma\in\{-,+\}} \frac{B^{(\overline{\sigma})}_{\sigma}(Q,\omega)}{J_\sigma(Q,\omega)}\mathbb{I}_{(Q,\omega)\in C_{\sigma}}
\end{equation}
or more explicitly
\begin{eqnarray}
S^{-+}_{(2)}(Q,\omega)=\frac{1}{2}\frac{1}{\omega T}\sum_{\sigma\in\{-,+\}}\frac{1+\sigma\cos(Q)}{W_{\sigma}}
\frac{\vartheta_{A}^2(\beta^{(\sigma)}_{-})}{\vartheta_{d}^2(\beta^{(\sigma)}_{-})} \times \nonumber \\
\times \left[\frac{1-\kappa}{1+\kappa}[\omega^2+\kappa\omega_0^2+T]\delta_{\sigma,+}+[\omega^2-\kappa\omega_0^2+T]\delta_{\sigma,-}\right]
\mathbb{I}_{(Q,\omega)\in C_{\sigma}}
\label{final}
\end{eqnarray}
with $\mathbb{I}_{(Q,\omega)\in C_{\sigma}}$ being one if $(Q,\omega)$ is within the region $C_{\sigma}$. 
Note that this result does not agree with the one in \cite{BougourziPRB57}, since there is only one weight
per sheet and not the sum of both weights.  Since the two weights are different, the TDSF is asymmetric around $Q=\pi/2$
for any value of anisotropy $\Delta < \infty$ away from the pure Ising limit.

We can also see how in this case the result in the isotropic point naturally arises. 
Indeed, as we can see from eq. (\ref{final}) the isotropic limit $\Delta \to 1$ corresponds to $\kappa\to 1$, 
which implies that the weight in the sheet $C_{+}$ vanishes and only the weighted sheet $C_{-}$ remains. 
Conservely, in the Ising limit $\kappa\to 0$, this asymmetry disappears.  This is therefore in complete correspondence
with what we have described earlier for the finite lattice.

\subsection{Results}
\begin{figure}[h]
\psfrag{omega}{\mbox{$\omega$}}
\psfrag{Q}{\mbox{$Q$}}
\includegraphics[height=7cm, width=7cm ]{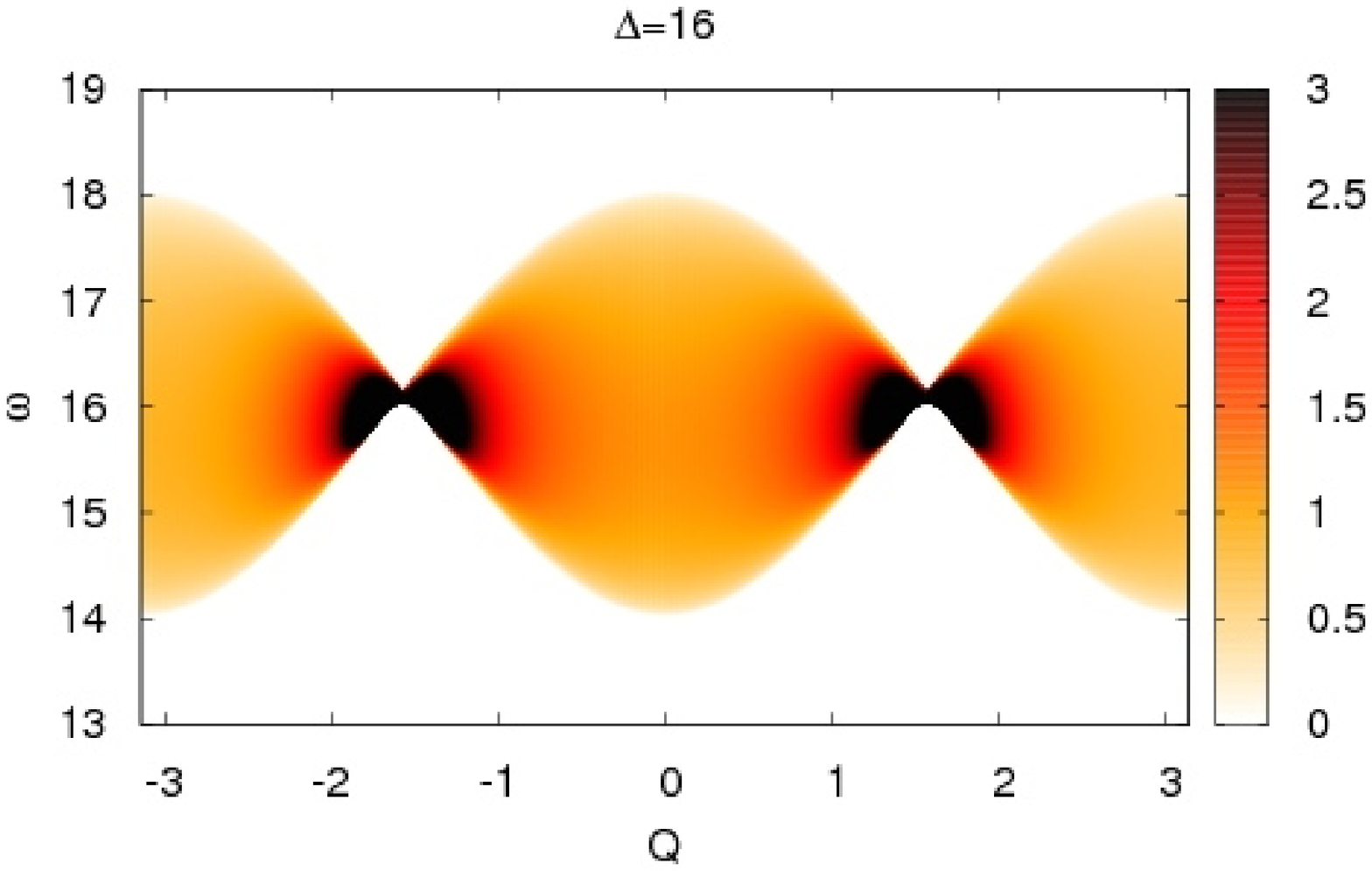}
\includegraphics[height=7cm, width=7cm ]{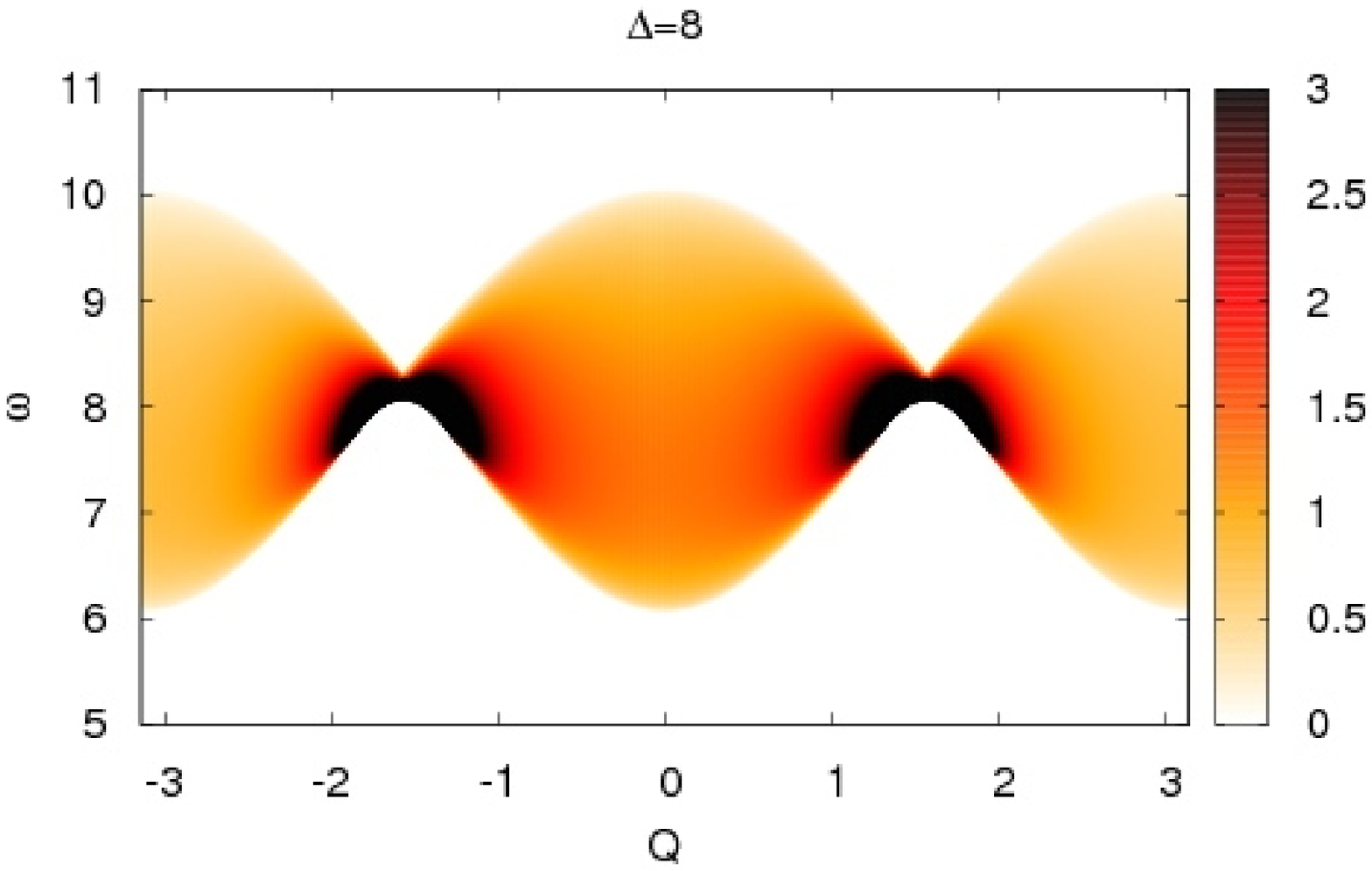} \\
\includegraphics[height=7cm, width=7cm ]{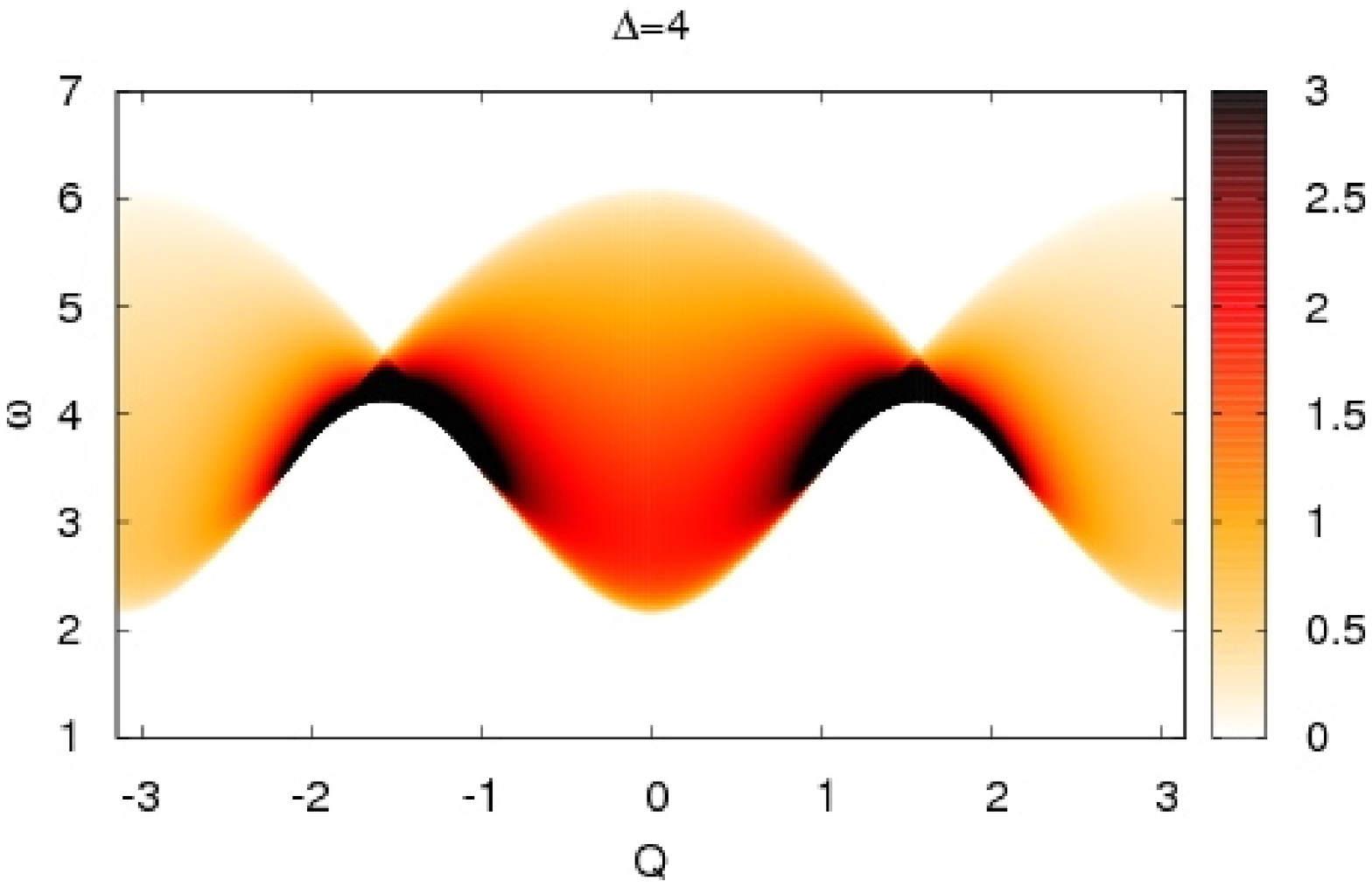}
\includegraphics[height=7cm, width=7 cm ]{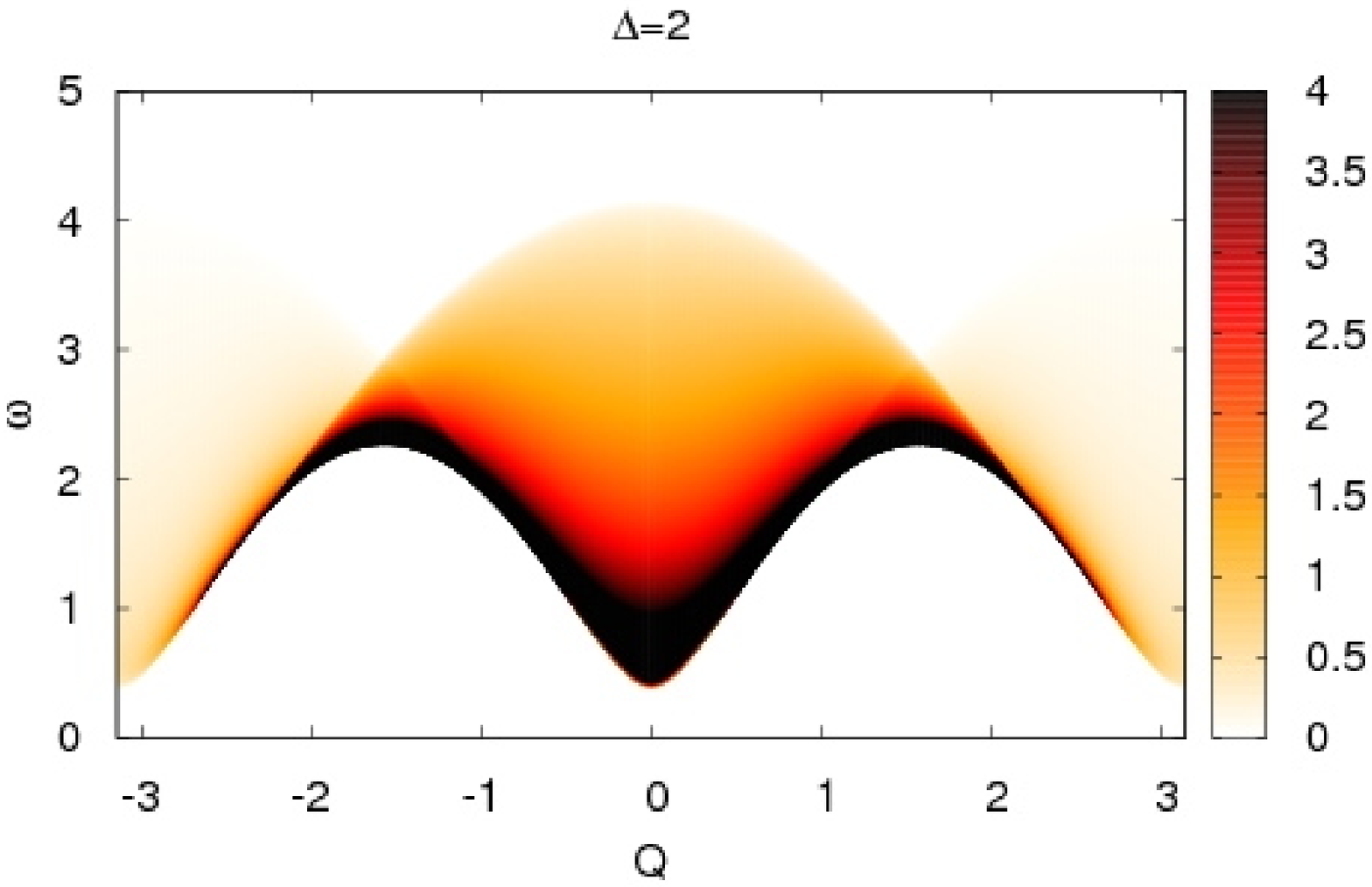}
\caption{Two-spinon transverse dynamical structure factor at zero field
for $\Delta=16$, $\Delta=8$, $\Delta=4$ and $\Delta=2$.  The approach towards the isotropic
limit is clearly seen, featuring vanishing of the gap and of the structure factor at the zone edges.
The asymmetry around the $\pi/2$ line is clearly seen to disappear only in the Ising limit $\Delta \rightarrow \infty$.}
\label{DSF1}
\end{figure}

\begin{figure}
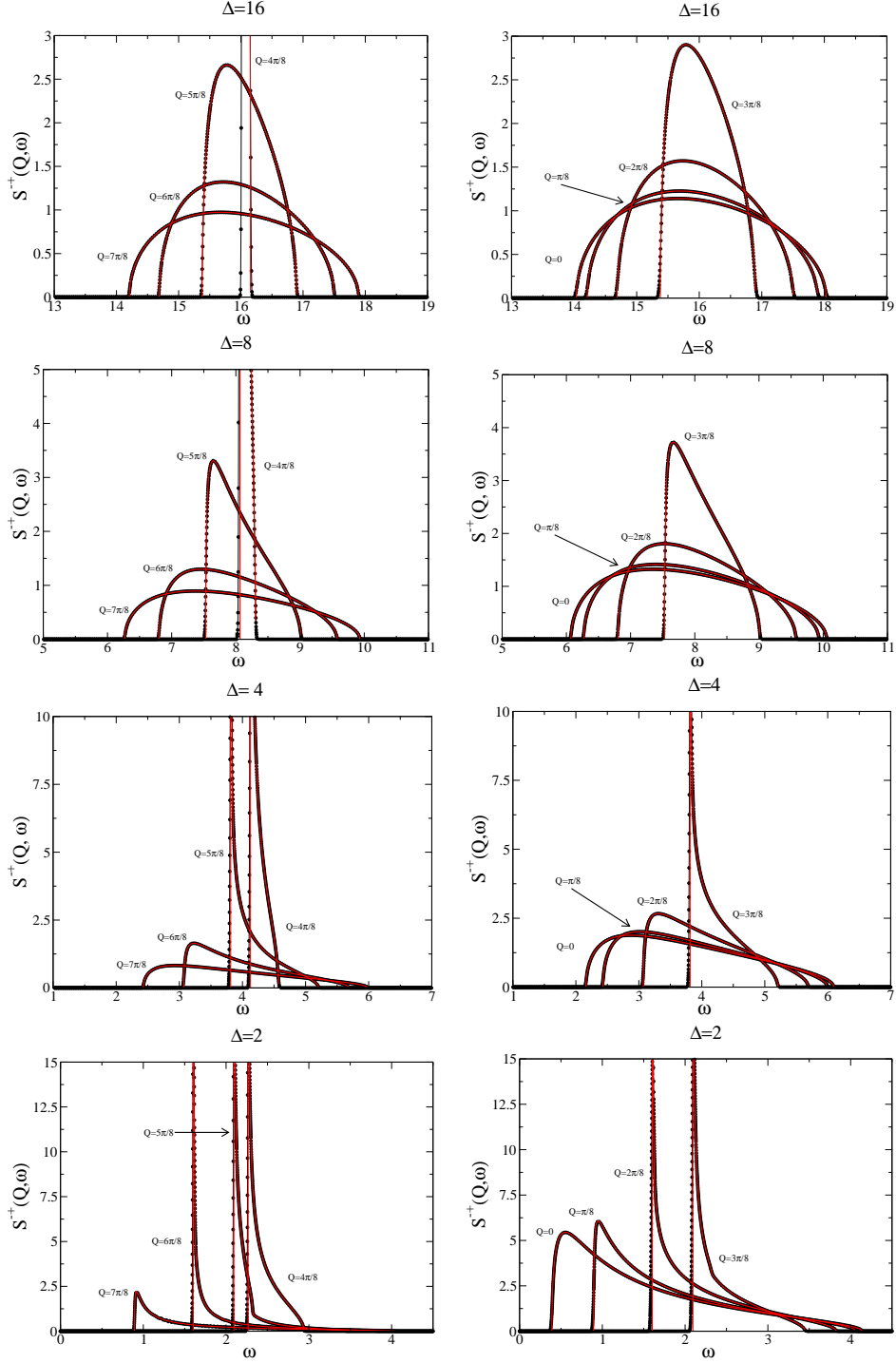

\begin{tabular}{cc}
\includegraphics[width=6cm]{XXZ_gpd_2spinons_Fig_5a.eps} & 
\includegraphics[width=6cm]{XXZ_gpd_2spinons_Fig_5b.eps} \\
\includegraphics[width=6cm]{XXZ_gpd_2spinons_Fig_5c.eps} &
\includegraphics[width=6cm]{XXZ_gpd_2spinons_Fig_5d.eps} \\
\includegraphics[width=6cm]{XXZ_gpd_2spinons_Fig_5e.eps} &
\includegraphics[width=6cm]{XXZ_gpd_2spinons_Fig_5f.eps} \\
\includegraphics[width=6cm]{XXZ_gpd_2spinons_Fig_5g.eps} &
\includegraphics[width=6cm]{XXZ_gpd_2spinons_Fig_5h.eps}
\end{tabular}
\caption{Fixed momentum cuts of the two-spinon transverse dynamical structure factor at zero field, for the same anisotropy
values as in Figure \ref{DSF1}.  The continuous lines are obtained from the algebraic analysis expressions,
and the points are obtained by a smoothing of the finite lattice result for $N = 1600$ sites.  }
\label{DSF2}
\end{figure}

Let us now present our results for the two-spinon part of the zero-field TDSF in the gapped antiferromagnetic regime.
In Fig. \ref{DSF1}, we plot the TDSF over all values of momentum and energy covered by the two-spinon
states, for four values of the anisotropy parameter.
As can be seen, by varying $\Delta$ we smoothly go from the Ising-like limit to the one of the isotropic point. 
The finite size results are not plotted here, since they are essentially identical to the ones presented.
Figure \ref{DSF2} provides a set of fixed-momentum cuts for the same four values of anisotropy, this time
showing both the infinite size (solid line) and finite size (colored points, computed for a lattice of $N = 1600$ sites) results.
The accurate agreement between the two approaches demonstrates that we have put the building blocks together in the right way.
As explained in \cite{BougourziPRB57}, the TDSF is characterized by square-root cusps at the lower and upper thresholds of
the two-spinon continuum, except for $Q$ within the range $Q_{\kappa}, \pi - Q_{\kappa}$, where the structure factor 
obtains a square root divergence at the lower threshold (the latter being given by $\omega_0(Q)$).
At the isotropic point, the divergence at the lower threshold covers the whole momentum
interval.  Note that our results are however different from those in \cite{BougourziPRB57}, for the reasons explained above.

The TDSF obeys a number of sum rules, two of which are of particular importance for our purposes.
First of all, the total integrated intensity is such that
\begin{eqnarray}
\frac{1}{N} \sum_k \int_{-\infty}^{\infty} \frac{d\omega}{2\pi} S^{-+}(k, \omega) = \frac{1}{2}.
\label{SR0}
\end{eqnarray}
Perhaps more importantly, the first frequency moment at fixed momentum obeys the sum rule
\cite{HohenbergPRB10}
\footnote{Remember that we have used different definitions of the Hamiltonian for the finite 
and infinite lattices.  In the structure factor expressions, $k$ on the finite lattice 
is thus equivalent to $\pi - Q$ on the infinite one.}
\begin{eqnarray}
\int_{-\infty}^{\infty} \frac{d\omega}{2\pi} \omega S^{-+} (k, \omega) 
= -\frac{2J}{N} \left[ (1 - \Delta \cos k) X^y + (\Delta - \cos k) X^z \right]
\label{SR1}
\end{eqnarray}
where $X^a \equiv \sum_j \langle S^a_j S^a_{j+1} \rangle$, $a = x, y, z$ are the expectation values of
the exchange terms.  Since these appear in the Hamiltonian, the value of the right-hand side
of (\ref{SR1}) is easily computed from the ground-state energy and its anisotropy dependence.

\begin{figure}
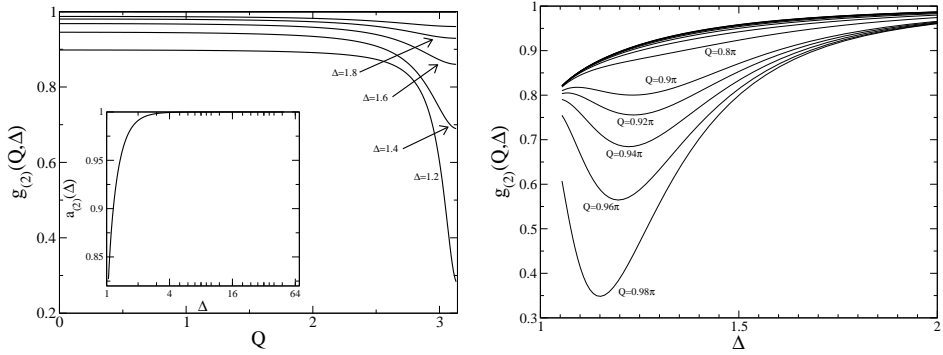

\begin{tabular}{cc}
\includegraphics[width=6cm]{XXZ_gpd_2spinons_Fig_6a.eps}
&
\includegraphics[width=6cm]{XXZ_gpd_2spinons_Fig_6b.eps}
\end{tabular}
\caption{Left (inset):  fraction of the total integrated intensity sum rule carried by two-spinon states, as a function
of anisotropy.  Left, main figure:  first moment sum rule fraction carried by two-spinon states, 
as a function of momentum, for different values of anisotropy.  Right:  anisotropy
dependence of the first moment sum rule fraction for fixed values of momentum.}
\label{fig:sumrules}
\end{figure}

In the Ising limit, two-spinon states completely saturate both sum rules.  In the isotropic limit,
it is known \cite{KarbachPRB55} that two-spinon states carry $72.89\%$ of the total integrated
intensity and $71.30\%$ of the first frequency moment sum rule.  
Figure \ref{fig:sumrules} presents our results for the contribution of two-spinon states to
both of these sum rules.  The quantity $a_2 (\Delta)$ is defined as the fraction of the total
integrated intensity carried by two-spinon states, and similarly $g_2 (Q, \Delta)$ is the
fraction of the first frequency moment at fixed momentum carried by two-spinon states.

The total integrated intensity quickly becomes saturated to high accuracy when $\Delta$
goes deeper in the gapped regime.  This is plotted in the first panel of Figure \ref{fig:sumrules}.
A more interesting point is that $g_2 (Q, \Delta = 1)$ does not depend on momentum \cite{KarbachPRB55},
but develops such a dependence away from the isotropic point.  This is illustrated in the second
panel of the Figure \ref{fig:sumrules}.  Most of the momentum dependence occurs close to the $Q = \pi$ point:
in the isotropic limit, the right-hand side of (\ref{SR1}) vanishes for $Q = \pi$ (and so does the
TDSF), but not for $\Delta > 1$.  As a function of $\Delta$, an interesting non-monotonic structure
is seen as a function of momentum, which is illustrated in the third panel of Figure \ref{fig:sumrules}.
The two-spinon intermediate states clearly carry the bulk of the TDSF in this regime.

\section{Conclusions and perspectives}
The understanding of the dynamics of strongly-correlated systems is clearly one of the
most challenging and long-standing problems in condensed matter, and integrable models
now provide a pathway towards achieving this goal.  Interestingly, two independent treatments
can be offered in the zero field chain case we considered, 
based either on integrability of the finite lattice, or on the quantum
group symmetry of the infinite chain.  Both approaches have their advantages and disadvantages:
the former is applicable to generic chains at generic magnetic fields, but is restricted to 
finite chains.  The latter is only valid for zero field, and cannot be used to understand
finite size effects.  Together, however, the two approaches paint a rather complete picture for zero field.

In future publications, we will consider the longitudinal structure factor $S^{zz}(k, \omega)$ in
the same regime, as well as the four-spinon contribution to the transverse structure factor, thereby
generalizing the recent results of the isotropic case \cite{CauxJSTATP12013} to this sector.  An
interesting further line of investigation would be to apply the algebraic approach within the gapless
anisotropic regime.  Finite temperature results would also be of great interest.

\section*{Acknowledgments}
J.-S. C. acknowledges support from the Stichting voor Fundamenteel Onderzoek der Materie 
(FOM) in the Netherlands.  We thank F. H. L. Essler, A. James, M. Jimbo and R. Weston together
with M. Karbach and G. M\"uller for useful discussions.

\vspace{1cm}

\bibliographystyle{jpa}
\bibliography{XXZ_gpd_2spinons}

\end{document}